\documentclass[12pt]{article}
\usepackage{setspace}
\usepackage{epsfig}
\usepackage{amsmath}
\usepackage{amssymb}
\usepackage{graphicx}
\def\be{\begin{equation}}
\def\ee{\end{equation}}
\def\ba{\begin{array}}
\def\ea{\end{array}}
\def\beqn{\begin{eqnarray}}
\def\eeqn{\end{eqnarray}}

\def\bt{\begin{tabular}}
\def\et{\end{tabular}}
\def\bc{\begin{center}}
\def\ec{\end{center}}

\begin{document}
\title{Fermion mass matrices, textures and beyond}
\author{Manmohan Gupta$^*$, Priyanka Fakay, Samandeep Sharma$^@$, Gulsheen Ahuja \\
{\it Department of Physics, Panjab University,
 Chandigarh, India.}\\
{\it $^@$Department of Physics, GGDSD College, Chandigarh, India.}
\\{\it $^*$mmgupta@pu.ac.in}}

\maketitle

\begin{abstract}
The issue of texture specific fermion mass matrices have been
examined briefly from the `bottom-up' perspective. In case no
conditions are imposed, the texture {\it ans\"{a}tze} leads to a
large number of viable possibilities. However, besides textures,
if in case one incorporates the ideas of `natural mass matrices'
and uses the facility of Weak Basis Transformations, then one is
able to arrive at a minimal finite set of viable mass matrices in
the case of quarks.
\end{abstract}

Understanding fermion masses and mixings is one of the biggest
challenges in the present day High Energy Physics. One of the key
difficulties in this area is the fact that the fermion masses and
mixings span several orders of magnitude. In the case of charged
fermions, the range of masses is from $10^{5}$ eV to $ 10^{12}$
eV, corresponding respectively to the electron mass and the mass
of the top quark. Further, the absolute masses of the neutrinos
are not known, however, two of the lightest neutrino masses can be
of the order of a fraction of an eV, with no lower limit for the
third neutrino mass. In case the theory requires the existence of
right handed neutrinos, responsible for see-saw mechanism
\cite{seesaw}-\cite{seesaw5} with the mass range of
$10^{12}-10^{15}$ GeV, the fermion masses would then cover almost
25 orders of magnitude.

The problem gets further complicated when one notices that the
pattern  of mixings are also quite different in case of quarks and
leptons. In fact, in the case of quarks we have clearly
hierarchical structure  of the  mixing angles, for example, $
s_{12} \sim 0.22, s_{23} \sim 0.04, s_{13} \sim 0.004$. In
contrast, the two of the mixing angles in case of neutrinos are
quite large, whereas the third angle although small as compared to
the other two  angles yet it is of the order of the Cabibbo angle.
Similarly, the pattern of masses in the case of charged leptons
has a very well defined hierarchy, whereas in the case of neutrino
we may have normal/inverted hierarchy or degenerate scenario of
neutrino masses. Since the mixing matrices are related to the
corresponding mass matrices therefore formulating viable fermion
mass matrices becomes all the more complicated.

In the absence of fundamental theory of flavor physics wherein
fermion masses and mixings can be understood, the present day
phenomenological approaches can be broadly categorized as
`top-down' and `bottom-up'. The top-down approach essentially
starts with the formulation of mass matrices at the GUT scale,
whereas, the bottom-up approach starts with the phenomenological
mass matrices at the weak scale.  Despite large number of attempts
from the top-down perspective \cite{chen} yet we are not in a
position to incorporate the vast amount of data related to fermion
mixing within a consistent framework. In this context, therefore,
it is desirable to look at bottom-up approach
\cite{fritzsch}-\cite{singreview} consisting of finding the
phenomenological fermion mass matrices which are in tune with the
low energy data, i.e., observables like quark and lepton masses,
mixing angles in both the sectors, angles of the unitarity
triangle in the quark sector, etc.. Also, successful
phenomenological formulation of mass matrices may provide clues
for appropriate dynamical models, in particular, important clues
for their formulation at the GUT scale.

The purpose of the present work is to explore the essentials, from
a `bottom-up' approach perspective, needed to arrive at a minimal
set of fermion matrices which are compatible with the latest
mixing data. To this end, we have not gone into a detailed and
comprehensive analysis rather would like to present a brief
overview related to the issue mentioned above. Further, we would
like to discuss the possibility of arriving at a minimal set of
viable mass matrices using textures and other ideas.

To begin with, we discuss the earliest {\it ans\"{a}tz} made in
the context of quark mass matrices. The first step in this
direction was taken by Fritzsch \cite{frzans,frzans1}, essentially
laying down the path for future investigations in this direction.
According to his hypothesis, the $3\times 3$ mass matrices for the
up and down sectors, $M_U$ and $M_D$, are hermitian and are given
by \be M_U= \left( \ba  {ccc} 0 & A_U & 0
\\ A_U^{*} & 0 & B_U
\\
0 & B_U^{*} & C_U \ea \right), \qquad M_D= \left( \ba  {ccc} 0 &
A_D & 0 \\ A_D^{*} & 0 & B_D \\
 0 & B_D^{*} & C_D \ea \right) \,. \label{frians} \ee
Another {\it ans\"{a}tz} proposed by Stech \cite{stech} has the
following form for the mass matrices in the up and down sectors
\be M_U=S\,,\qquad M_D = \beta S + A \,,\ee where $S$ and $A$ are
symmetric and antisymmetric $3\times 3$ matrices respectively. Yet
another {\it ans\"{a}tz}, proposed by Gronau \cite{gronau1}, had
the features of both Fritzsch's and Stech's {\it ans\"{a}tze},
e.g, \be M_U= \left( \ba  {ccc} 0 & A & 0 \\ A & 0 & B \\
            0 & B & C \ea \right),  \qquad
M_D= \beta\left( \ba  {ccc} 0 & A & 0 \\ A & 0 & B \\
            0 & B & C \ea \right) +
\left( \ba  {ccc} 0 & ia& 0 \\-ia& 0 & ib\\
            0 &-ib& 0 \ea \right)
 \,. \ee

Interestingly, these {\it ans\"{a}tze} were ruled out by the
``high'' value of the $t$ quark mass and these continue to be
ruled out even with subsequent refinements in the data. To this
end, we discuss, in somewhat detail, the case of Fritzsch {\it
ans\"{a}tz}. The essentials of the methodology usually used to
carry out the analysis include diagonalizing the mass matrices
$M_U$ and $M_D$ by unitary transformations and obtaining a
Cabibbo-Kobayashi-Maskawa (CKM) matrix from these transformations.
To ensure the viability of the considered mass matrices, this CKM
matrix should be compatible with the quark mixing data, for
details regarding this we refer the readers to \cite{singreview}.
Following this methodology for the above mentioned {\it
ans\"{a}tz} considered by Fritzsch, the CKM matrix so obtained by
considering latest inputs from PDG 2014 \cite{pdg} is given by
 \be V_{{\rm CKM}} = \left( \ba{ccc}
  0.9837-0.9872 &~~~~   0.2248-0.2268 &~~~~  0.0053-0.0075 \\
 0.2203-0.2264  &~~~~   0.9160-0.9721    &~~~~  0.0601-0.2037\\
0.0302-0.0308  &~~~~  0.0043-0.0194 &~~~~  0.9991-0.9999
\label{ckmmatrix} \ea \right).  \ee A look at this matrix
immediately reveals that the ranges of most of the CKM elements
show no overlap with those obtained by recent global analyses
\cite{pdg}. This, therefore, leads to the conclusion that the
Fritzsch {\it ans\"{a}tz} is not compatible with the recent quark
mixing data.

The above conclusion can be explicitly understood by studying the
analytical expressions of the elements $|V_{ub}|$ and $|V_{cb}|$,
e.g.,
\be
V_{ub}=-\sqrt{\frac{m_d}{m_s}}({\frac{m_s}{m_d}})^{\frac{3}{2}}
e^{i\phi_1}-\sqrt{\frac{m_u}{m_c}}\sqrt{\frac{m_s}{m_b}}+
\sqrt{\frac{m_u}{m_c}}\sqrt{\frac{m_c}{m_t}}e^{i\phi_2},\ee

\be
V_{cb}=\sqrt{\frac{m_u}{m_c}}\sqrt{\frac{m_d}{m_s}}({\frac{m_s}{m_b}})^{\frac{3}{2}}
e^{i\phi_1}-\sqrt{\frac{m_s}{m_b}}+
\sqrt{\frac{m_c}{m_t}}e^{i\phi_2}, \ee where phases $\phi_1$ and
$\phi_2$ are related to the phases associated with the elements of
the mass matrices \cite{minireview}. In Fig.1, we have plotted the
dependence of these elements with respect to the strange quark
mass $m_s$. While plotting the allowed ranges of the matrix
elements $|V_{ub}|$ and $|V_{cb}|$, all other parameters have been
given full variation within the allowed ranges. A general look at
the figure immediately shows that the plotted values both
$|V_{ub}|$ and $|V_{cb}|$ have no overlap with the allowed
experimental ranges of these. Thus, one can again conclude that
Fritzsch {\it ans\"{a}tz} is not viable.

\begin{figure}[hbt]
\begin{minipage}{0.45\linewidth}   \centering
\includegraphics[width=2.in,angle=-90]{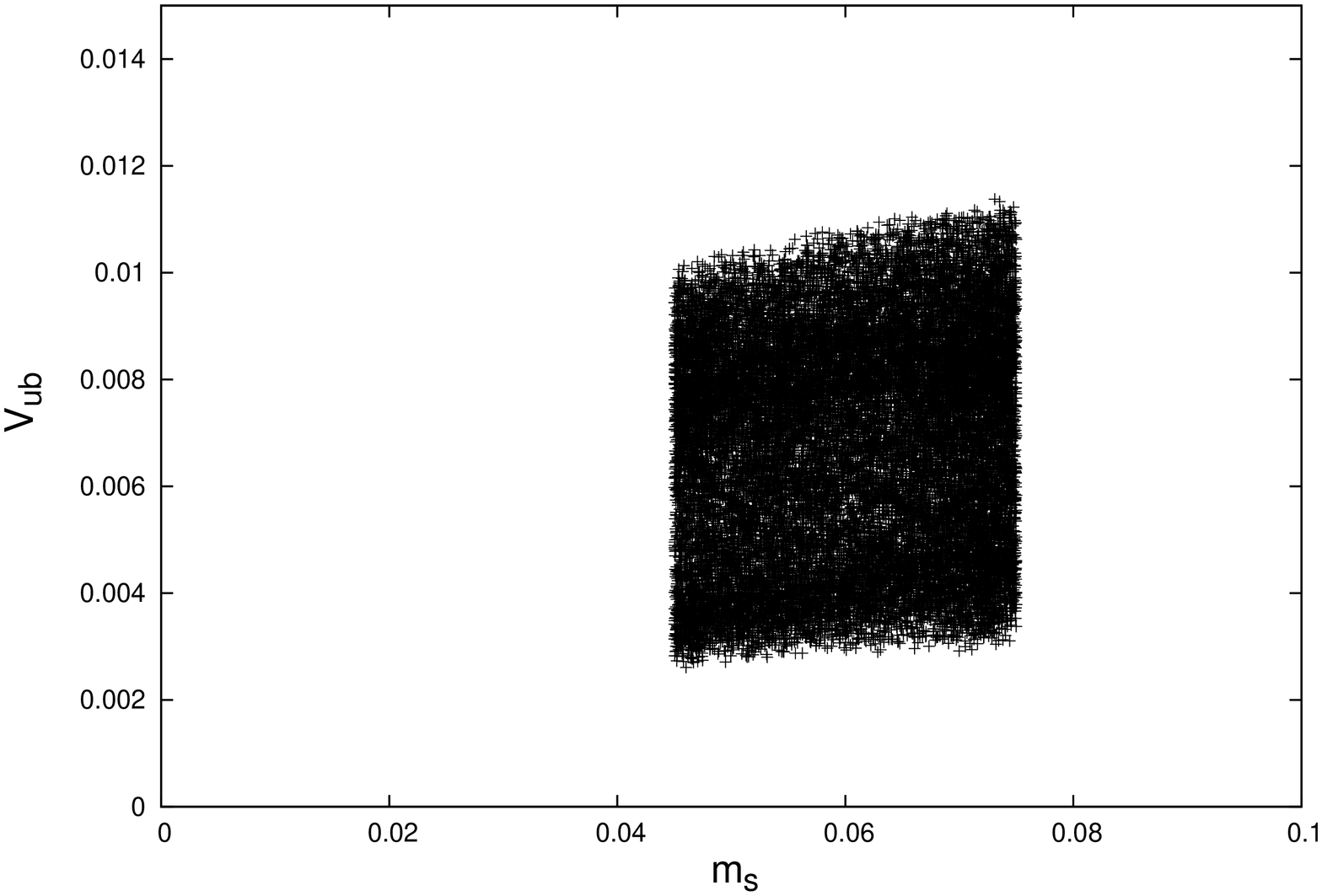}
\end{minipage} \hspace{0.5cm}
\begin{minipage} {0.45\linewidth} \centering
\includegraphics[width=2.in,angle=-90]{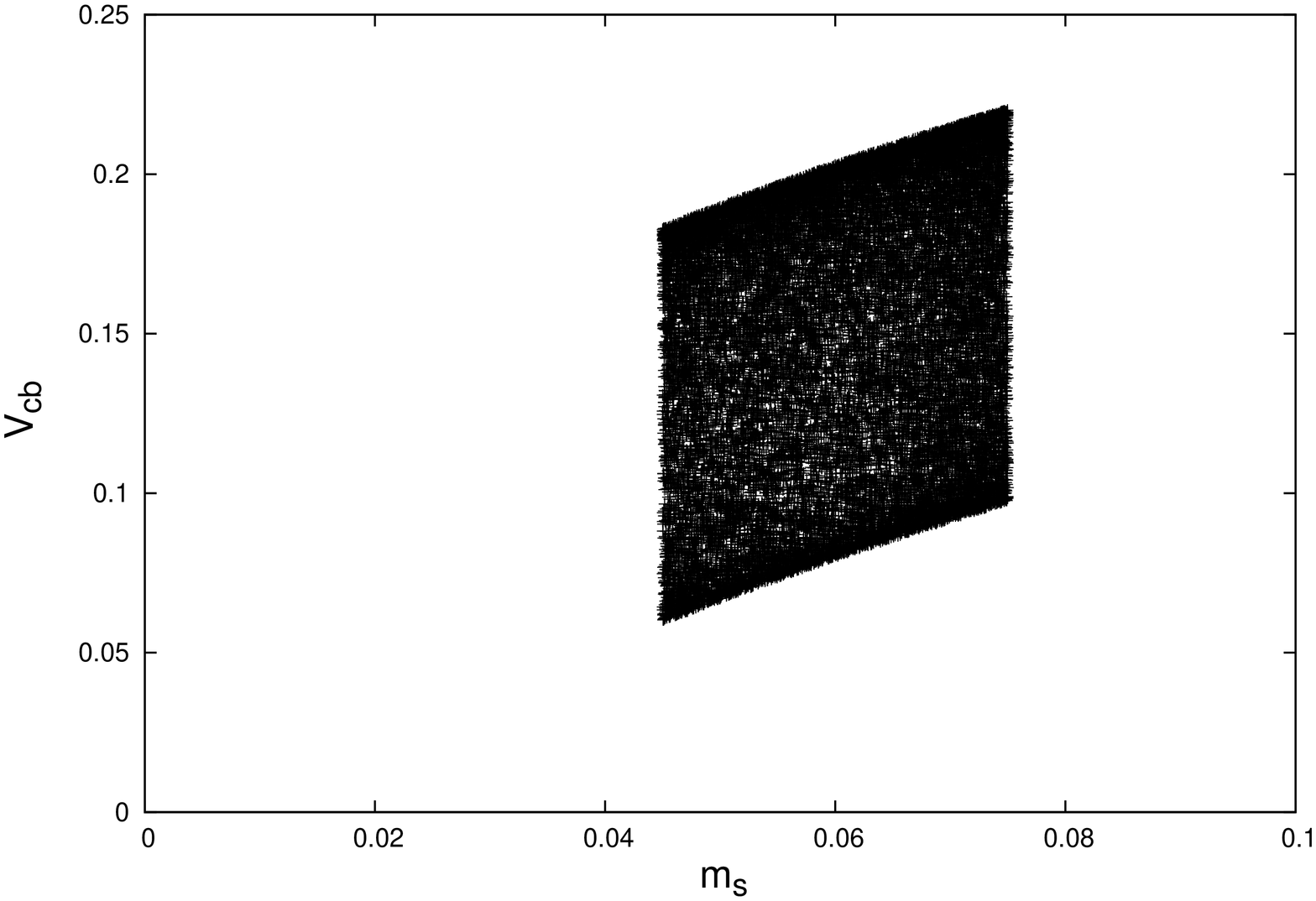}
\end{minipage}\hspace{0.5cm}
\caption{Plots showing the allowed range of $|V_{ub}|$ and
$|V_{cb}|$ w.r.t the light quark mass $m_s$ for the Fritzsch mass
matrix. } \label{nhih6z}
\end{figure}

The generalization of the Fritzsch {\it ans\"{a}tze} led to the
idea of textures. {\it A particular texture structure is said to
be texture $n$ zero, if it has $n$ number of non-trivial zeros,
for example, if the sum of the number  of diagonal zeros and half
the number of the symmetrically placed off diagonal zeros is $n$}.
Therefore, if both $M_U$ and $M_D$ have $n$ texture zeros each,
together these are called texture $2n$ zero mass matrices. For
example, the Fritzsch {\it ans\"{a}tz}, mentioned in equation
(\ref{frians}), corresponds to texture 6 zero quark mass matrices.

Apart from texture 6 zero mass matrices considered by Fritzsch,
some other versions of these were also analyzed and consequently
ruled out by  Ramond $\it {et.al.}$ \cite{rrr}, these continue to
be ruled out even by the present quark mixing data. In this
context, Ramond $\it {et.al.}$ \cite{rrr} also arrived at an
important conclusion that the texture structure of a matrix as
well as its hermiticity property are not `affected' when one
scales down from GUT scale to weak scale, justifying the
formulation of texture specific mass matrices. This important
conclusion also leads to the fact that the texture zeros of
fermion mass matrices can be considered as phenomenological zeros,
thereby implying that at all energy scales the corresponding
matrix elements are sufficiently suppressed in comparison with
their neighboring counterparts. This, therefore, opens the
possibility of considering less than six texture zeros
\cite{minireview} for the quark mass matrices.

Extending their analysis of texture 6 zero mass matrices, Ramond
$\it {et.al.}$ \cite{rrr} have also examined the viability of a
few texture 5 zero quark mass matrices. Recently, the
compatibility of texture 5 zero mass matrices with the latest
mixing data has also been examined in detail \cite{minireview}.
Interestingly, even in this case one finds that there is only
marginal compatibility, in particular, out of the large number of
possibilities for texture 5 zero mass matrices, only Fritzsch-like
mass matrices have limited compatibility with the experimental
data. As an extension of texture 5 zero mass matrices, several
authors have carried out the study of the implications of the
Fritzsch-like texture 4 zero mass matrices
\cite{xing4}-\cite{monika}. These analyses reveal that the texture
4 zero mass matrices, undoubtedly, are able to accommodate the
quark mixing data quite well.

Very recently, Ludl and Grimus \cite{ludl} have performed a
detailed and comprehensive analysis for general as well as
symmetric texture specific quark mass matrices. Without imposing
any restrictions on textures and using the facility of `Weak Basis
Transformations', Ludl and Grimus arrive at 243 classes of texture
specific mass matrices, related through permutations. To reduce
the number of possibilities they use the concept of maximally
restrictive classes (one cannot place another texture zero into
one of the two mass matrices while keeping the model compatible
with the data). Thus, they found 27 viable classes for general
mass matrices, however, without any predictive powers. In the case
of symmetric mass matrices they have found 15 maximally
restrictive textures which are predictive with respect to one or
more light quark masses.

The above analysis indicates that in the absence of any additional
conditions on textures, even texture 5 zero mass matrices could
also be viable and the number of viable possibilities increases
rapidly as one goes to lower textures. This therefore, brings us
to the conclusion that in case we have to arrive at finite set of
mass matrices which may serve as clues for their formulation at
fundamental level, one needs to go beyond texture {\it
ans\"{a}tze}. In this context, two important ideas for the quark
matrices have been considered in the literature, e.g., the concept
of `natural mass matrices', advocated by Peccei and Wang
\cite{nmm} and that of Weak Basis (WB) transformations, considered
by Fritzsch and Xing \cite{xingwb} as well as Branco {\it et.al.}
\cite{branco}.

The essential idea of `natural mass matrices' consists of
formulating quark mass matrices which are able to reproduce
hierarchical mixing angles without resorting to fine tuning. This
results in considerably constraining the parameter space available
to the elements of the mass matrices. Using this concept Peccei
and Wang \cite{nmm} have attempted to reconstruct mass matrices at
$M_z$ as well as GUT scale, however without invoking any other
condition they are not able to find any finite or viable set of
mass matrices. In the context of texture specific mass matrices,
the idea of `natural mass matrices' has been found to be useful in
reproducing the data when the following hierarchy is imposed on
the elements of the quark mass matrices
 \be (1,i)  \leq  (2,j)
\leq  (3,3) ~~~i=1,2,3 ; j=2,3. \ee

As mentioned earlier, Weak Basis transformations is an another
idea to go beyond texture {\it ans\"{a}tze}, considered by by
Fritzsch and Xing \cite{xingwb} as well as Branco {\it et.al.}
\cite{branco}. Within the framework of the SM, the hermitian quark
mass matrices, which encode all the information about the quark
masses and mixings, have a total of 18 real free parameters, which
is a large number compared to only ten physical parameters
corresponding to six quark masses and four physical parameters of
the CKM matrix. In this context, it is interesting to note that
one has the freedom to make a unitary transformation, e.g.,
 $ q_L\rightarrow W q_L~,~ q_R \rightarrow W q_R $,~~ $
q_L^{\prime}\rightarrow W q_L ^{\prime}~,~ q_R^{\prime}
\rightarrow W q_R ^\prime $ under which the gauge currents \beqn -
{\cal L}^{cc}_{W} & = & \frac{g}{\sqrt{2}} \overline {(u,c,t)}_{L}
\gamma^{\mu}\left( \ba{c} d \\ s\\ b \ea \right)_L\ W_{\mu} +
h.c.\,\label{ckm} \eeqn
 remain real and diagonal but the mass matrices
transform as  \beqn M_U \longrightarrow M_U^\prime =
W^{\dagger}M_{U}W,~~ M_D \longrightarrow M_D^\prime =
W^{\dagger}M_{D}W \eeqn where W is an arbitrary unitary matrix.
Such transformations are referred to as `Weak Basis (WB)
Transformations'.

The WB transformations broadly lead to two possibilities for the
texture zero fermion mass matrices. In the first possibility, as
observed by Fritzsch and Xing \cite{xingwb}, one ends up with
texture 2 zero fermion mass matrices, wherein both the fermion
mass matrices assume a texture 1 zero hermitian structure of the
following form
\be
M_q^{'}=\left( \ba {ccc} {*} & * & 0 \\ {*} & {*} & {*} \\ 0 & * &
* \ea \right), ~~~~~~q^{'}=U,D.
\ee In the second possibility, as observed by Branco {\it{et al.}}
\cite{branco} one ends up with texture 3 zero fermion mass
matrices $M_U$ and $M_D$ wherein one of the matrix among these
pairs is a texture 2 zero Fritzsch-like hermitian mass matrix
given by
\be
M_q=\left( \ba {ccc} 0 & * & 0 \\
                     {*} & * & * \\
                     0 & * & * \ea \right), ~~~~~~q=U,D,
                   \ee
while the other mass matrix is a texture 1 zero hermitian mass
matrix of the following form
\be
M_q^{'}=\left( \ba {ccc} 0 & * & * \\
                     {*} & * & * \\
                     {*} & * & * \ea \right), ~~~~~~q^{'}=U,D.
                   \ee
Further, we would like to emphasize here that although the two
approaches for WB transformations are equivalent, but the approach
by Branco {\it{et al.}} leads to non parallel texture three zero
structure while the approach  by Fritzsch and Xing leads to
parallel texture two zero structure.

Recently an analysis by Costa and Simoes \cite{costa} shows that
starting from arbitrary matrices $M_U$ and $M_D$, it is always
possible to perform a WB transformation that renders them
Hermitian with a particular texture, therefore, resulting in
reducing the number of free parameters of general mass matrices.
The obtained quark matrices are confronted with the experimental
data, reconstructing them at the electroweak scale and at a high
scale where the Froggatt-Nielsen mechanism can be implemented.
However, in the absence of any constraints on the elements of the
mass matrices, it leads to a large number of viable texture zero
matrices.

It is therefore evident from the above discussion that neither
texture {\it ans\"{a}tze} nor  Weak Basis transformations or
`naturalness' criteria, on their own, are able to lead to a finite
set of viable texture specific mass matrices. In order to obtain
the same, perhaps one needs to combine the three as discussed
recently by Sharma {\it{et al.}}\cite{prd}. This analysis shows
that one can start with the most general mass matrices and
consequently explore the possibility of obtaining a finite set of
viable texture specific mass matrices formulated by using weak
basis transformations as well as the constraints imposed due to
`naturalness'. Interestingly, the analysis reveals that a
particular set of texture 4 zero quark mass matrices can be
considered to be a unique viable option for the description of
quark mixing data.

A corresponding analysis in the lepton sector, wherein one
explores the possibility of arriving at a minimal set of lepton
texture specific mass matrices, reveals that this is not possible
because of a large number of viable possibilities. The analysis
pertaining to texture 4 zero Fritzsch-like mass matrices in the
Dirac as well as Majorana neutrino case indicates that these
matrices are compatible with the normal hierarchy and degenerate
scenario of neutrino masses whereas for inverted hierarchy such
matrices are ruled out in case the naturalness conditions are
imposed. In conclusion, we can perhaps say that the texture 4 zero
Fritzsch-like mass matrices provide an almost unique class of
viable fermion mass matrices giving vital clues towards unified
textures for model builders.

\vskip 0.5cm {\bf Acknowledgements} \\ M.G. and P.F.  would like
to acknowledge CSIR, Govt. of India, (Grant No:03:(1313)14/EMR-II)
for financial support. S.S. acknowledges the Principal, GGDSD
College, Sector 32, Chandigarh.  G.A. would like to acknowledge
DST, Government of India (Grant No: SR/FTP/PS-017/2012) for
financial support.  P.F. and S.S. acknowledge the Chairperson,
Department of Physics, P.U., for providing facilities to work.


\begin{thebibliography}{10}

\bibitem{seesaw}H. Fritzsch, M. Gell-Mann, P. Minkowski, Phys Lett.
{\bf B 59}, 256 (1975).

\bibitem{seesaw1}P. Minkowski, Phys. Lett. {\bf B 67}, 421 (1977).

\bibitem{seesaw2}T. Yanagida, in {\it Proceedings of the Workshop on Unified Theory
and the Baryon Number of the Universe}, edited by O. Sawada, A.
Sugamoto (KEK, Tsukuba, 1979), p. {\bf 95}.

\bibitem{seesaw3}M. Gell-Mann, P.
Ramond, R. Slansky, in {\it Supergravity}, edited by F. van
Nieuwenhuizen and D. Freedman (North Holland, Amsterdam, 1979), p.
{\bf 315}.

\bibitem{seesaw4}S. L. Glashow, in {\it Quarks and Leptons}, edited by M.
L$\rm\acute{e}$vy {\it et al.} (Plenum, New York, 1980), p. {\bf
707}.

\bibitem{seesaw5}R. N. Mohapatra, G. Senjanovic, Phys. Rev. Lett. {\bf 44},
912 (1980).

\bibitem{chen}Mu-Chen, K. T. Mahanthappa, Int. Jour. of Mod.
Phys. {\bf A 18}, 5819 (2003).

\bibitem{fritzsch}H. Fritzsch, Z.Z. Xing,  Nucl. Phys.  {\bf B 556},
49 (1999) and references therein.

\bibitem{xing1}Z. Z. Xing, H. Zhang, J. Phys. {\bf G 30},
129 (2004) and refrences therein.

\bibitem{minireview}M. Gupta, G. Ahuja, Int. J. Mod. Phys. {\bf A 26},
2973 (2011) and refrences therein.

\bibitem{singreview}M. Gupta, G. Ahuja, Int. J. Mod. Phys. {\bf A 27},
1230033 (2012) and references therein.

\bibitem{frzans}H. Fritzsch, Phys. Lett. {\bf B 70}, 436 (1977).

\bibitem{frzans1}H. Fritzsch, Phys. Lett. {\bf B 73}, 317 (1978).

\bibitem{stech}B. Stech, Phys. Lett. {\bf B 130}, 189 (1983).

\bibitem{gronau1}M. Gronau, R. Johnson, J. Schechter,
Phys. Rev. Lett. {\bf 54}, 2176 (1985).

\bibitem{rrr}P. Ramond, R.G. Roberts, G.G. Ross, Nucl. Phys. {\bf B 406}, 19
(1993).

\bibitem{pdg}K. A. Olive $\it {et.al.}$ Particle Data Group, Chin. Phys. {\bf  C 38}, 090001
(2014).

\bibitem{xing4}D. Du, Z. Z. Xing, Phys. Rev. {\bf D 48}, 2349 (1993).

\bibitem{pramana}P. S. Gill, M. Gupta, Pramana {\bf 45}, 333 (1995).

\bibitem{gill}P. S. Gill, M. Gupta, Phys. Rev. {\bf D 57}, 3971 (1998).

\bibitem{monika}M. Randhawa, M. Gupta,  Phys. Rev. {\bf D 63}, 097301 (2001).

\bibitem{ludl}P. Ludl, W. Grimus, arXiv: hep-ph/1501.04942.

\bibitem{nmm}R.D. Peccei, K. Wang, Phys. Rev. {\bf D 53}, 5 (1996).

\bibitem{xingwb}H. Fritzsch, Z. Z. Xing,  Phys. Lett.  {\bf B 413}, 396 (1997)
and references therein.

\bibitem{branco}G. C. Branco {\it et al.}, Phys. Rev. Lett. {\bf 82}, 683 (1999).

\bibitem{costa}D. Emmanuel-Costa, C. Simoes, Phys. Rev. {\bf D 79}, 073006 (2009).

\bibitem{prd}S.Sharma, P. Fakay, G. Ahuja, M. Gupta, Phys. Rev. {\bf D 91}, 053004 (2015).

\end{thebibliography}
\end{document}